\documentclass[nofootinbib,reprint,superscriptaddress,aps]{revtex4-2}
\usepackage{times}
\usepackage{amssymb}
\usepackage{color}
\usepackage{amsmath}
\usepackage{amsbsy}
\usepackage{amsthm}
\usepackage{graphicx}
\usepackage{bbm}
\usepackage{bm}
\usepackage{epsfig}
\usepackage{xfrac}
\usepackage{xcolor}
\usepackage{enumerate}
\usepackage{multirow}
\usepackage{physics}

\usepackage[T1]{fontenc}
\usepackage[english]{babel}

\usepackage{appendix}
\usepackage{dsfont}
\usepackage{float}

\usepackage{relsize}
\usepackage{xcolor}
\usepackage{scalerel}
\usepackage{tikz}
\usetikzlibrary{svg.path}
\usepackage{ulem}

\definecolor{orcidlogocol}{HTML}{A6CE39}
\tikzset{
  orcidlogo/.pic={
    \fill[orcidlogocol] svg{M256,128c0,70.7-57.3,128-128,128C57.3,256,0,198.7,0,128C0,57.3,57.3,0,128,0C198.7,0,256,57.3,256,128z};
    \fill[white] svg{M86.3,186.2H70.9V79.1h15.4v48.4V186.2z}
                 svg{M108.9,79.1h41.6c39.6,0,57,28.3,57,53.6c0,27.5-21.5,53.6-56.8,53.6h-41.8V79.1z M124.3,172.4h24.5c34.9,0,42.9-26.5,42.9-39.7c0-21.5-13.7-39.7-43.7-39.7h-23.7V172.4z}
                 svg{M88.7,56.8c0,5.5-4.5,10.1-10.1,10.1c-5.6,0-10.1-4.6-10.1-10.1c0-5.6,4.5-10.1,10.1-10.1C84.2,46.7,88.7,51.3,88.7,56.8z};
  }
}

\newcommand\orcid[1]{\href{https://orcid.org/#1}{\mbox{\scalerel*{
\begin{tikzpicture}[yscale=-1,transform shape]
\pic{orcidlogo};
\end{tikzpicture}
}{|}}}}

\definecolor{myurlcolor}{rgb}{0,0,0.7}
\definecolor{myrefcolor}{rgb}{0.8,0,0}

\usepackage[colorlinks]{hyperref}
\hypersetup{unicode=true,
    bookmarksopen=false,
    breaklinks=false,
    pdfborder={0 0 0},
	bookmarksnumbered=false,
	pdfstartview={FitH},
	citecolor={myurlcolor},
	linkcolor={myrefcolor},
	urlcolor={myurlcolor}}

\definecolor{cyan(process)}{rgb}{0.0, 0.72, 0.92}

\usepackage{soul}
\renewcommand{\t}[1]{\textrm{#1}}

\begin{document}
\title{Universality of quantum time dilation}

\author{Kacper~D\k{e}bski\orcid{0000-0002-8865-9066}} \email{kdebski@fuw.edu.pl} \affiliation{Institute of Theoretical Physics, University of Warsaw, Pasteura 5, 02-093 Warsaw, Poland}

\author{Piotr~T.~Grochowski\orcid{0000-0002-9654-4824}} \email{piotr.grochowski@uibk.ac.at} \affiliation{Institute for Quantum Optics and Quantum Information of the Austrian Academy of Sciences, A-6020 Innsbruck, Austria} \affiliation{Institute for Theoretical Physics, University of Innsbruck, A-6020 Innsbruck, Austria} \affiliation{Center for Theoretical Physics, Polish Academy of Sciences, Aleja Lotnik\'ow 32/46, 02-668 Warsaw, Poland}
 
\author{Rafał~Demkowicz-Dobrzański\orcid{0000-0001-5550-4431}} \email{demko@fuw.edu.pl} \affiliation{Institute of Theoretical Physics, University of Warsaw, Pasteura 5, 02-093 Warsaw, Poland}
 
\author{Andrzej~Dragan\orcid{0000-0002-5254-710X}} \email{dragan@fuw.edu.pl} \affiliation{Institute of Theoretical Physics, University of Warsaw, Pasteura 5, 02-093 Warsaw, Poland} \affiliation{Centre for Quantum Technologies, National University of Singapore, 3 Science Drive 2, 117543 Singapore, Singapore}

\begin{abstract}
Time dilation is a difference in measured time between two clocks that either move with different velocities or experience different gravitational potentials. Both of these effects stem from the theory of relativity and are usually associated with classically defined trajectories, characterized by position, momentum, and acceleration. However, when spatial degrees of freedom are treated in a quantum way and a clock is allowed to be in a coherent superposition of either two momenta or two heights, additional quantum corrections to classical time dilation appear, called kinematic and gravitational quantum time dilations, respectively. We show that similarly to its classical counterpart, kinematic quantum time dilation is universal for any clock mechanism, while gravitational quantum time dilation is not. We also show that although both of these effects reduce to incoherent averaging of different classical time dilation contributions, there exists an additional quantum time dilation effect that has no classical analog and can be extracted from higher-order corrections to the system's Hamiltonian.
\end{abstract}

\maketitle 

\textbf{Introduction}---Pendulum clocks, atomic clocks, or any other clocks undergo a universal time dilation when moving at constant velocity. This universality is a direct consequence of the principle of relativity---motion at constant velocity is relative. It follows from this principle that a scenario with two different clocks that move together at the same velocity relative to some observer, is equivalent to a scenario with both clocks being at rest, but the observer moving in the opposite direction. In the latter case, both clocks must be affected the same way, because none of them is preferred in this setting and therefore the time dilation must be universal, i.e., independent of the clock mechanism.

However, for motion with nonuniform velocity, the above reasoning does not apply and therefore time dilation is not universal \cite{Dragan2021}.
For example, the effect of acceleration on the rate of moving clocks depends on their mechanism. A pendulum clock will react differently to acceleration than an atomic clock. Moreover, it was shown \cite{Lorek2015,EISELE1987,Pierini2018} that no physical system acting as a clock can be insensitive to accelerations due to the Unruh effect \cite{Birrell} and therefore no ideal clocks can exist. However, an idealized clock as a physical model is still commonly used and studied in physics. Additionally, one of the most promising approaches to understanding the nature of time involves relational formulation of quantum mechanics \cite{Rovelli1996}, including studying the properties of quantum clocks \cite{Vedral2007,Reinhardt2007,Zych2011,Bushev2016,Ruiz2017,Loriani2019,Smith2019,Paige2020,Smith2020arxiv,Roura2021}.

Recently, several authors have studied a scenario in which a quantum clock moves in a superposition of motions at different, but constant velocities, and a novel effect called quantum time dilation was introduced and studied \cite{Smith2020, Grochowski2021}. Similar effects were investigated in the studies of gravitational time dilation when a clock was placed in a superposition of two locations at different heights \cite{Khandelwal2020,Paczos2022,Hadi2022}. Consistency of the obtained results prompted the authors to speculate that quantum time dilation in analogy to its classical counterpart may also be universal. Indeed, we know that rates of clocks moving along classically defined inertial trajectories exhibit the same dilation regardless of their physical mechanism. The question remains whether the same universality applies to clocks following quantum-mechanical superpositions of inertial trajectories. In this case, the universality of quantum time dilation would be manifested as the independence of the effect of the Hilbert space of the clock, or the specifics of its Hamiltonian. Such an eventuality would lead to important consequences for the principle of relativity, which is closely related to the universality of time dilation. In this case, it would be possible to generalize this principle to scenarios, when inertial observers follow a superposition of paths \cite{Angelo2011,Kabel2022,Hamette}.

In this work, we provide the first proof of the universality of kinematic quantum time dilation and show that the effect must be the same for all clock mechanisms regardless of their Hamiltonian or Hilbert space. We also argue that gravitational time dilation cannot be universal for the same reason that time dilation along an accelerated trajectory cannot. Therefore the effect of gravity on the clock's rate must fundamentally depend on the clock's mechanism.

\textbf{Universality of kinematic quantum time dilation}---A relation between energy $E$ and momentum $\boldsymbol{p}$ of a classical pointlike particle of a mass $m$, placed at position $\boldsymbol{r}$ in a stationary gravitational field $g_{\mu\nu}$ with $g_{0i}=0$ is given by:
\begin{align}
E  = \sqrt{g_{00}(\boldsymbol{r})}\sqrt{\left(mc^2\right)^2 - g^{ij}(\boldsymbol{r})p_i p_j c^2}.
\label{eqn:Elab}
\end{align}
In order to quantize this classical scheme we consider a general quantum system characterized by the Hilbert space $\mathcal{H}_{\t{cm}} \otimes \mathcal{H}_{\t{clock}}$ consisting of center-of-mass degrees of freedom of the clock $\mathcal{H}_{\t{cm}}$, and its internal degrees of freedom $\mathcal{H}_{\t{clock}}$ representing the actual time measuring mechanism. The interaction between these degrees of freedom will result in a change in the pace of ``clock ticking'' due to motion or gravity. 

If the internal degrees of freedom are characterized by the Hamiltonian $\hat{H}_{\t{clock}}$ and the total Hamiltonian of the system is $\hat{H}$ then the first quantization procedure of the relation \eqref{eqn:Elab} leads to the total Hamiltonian:
\begin{align}
\hat{H}  = \sqrt{g_{00}(\hat{\boldsymbol{r}})}\sqrt{\left(mc^2 + \hat{H}_{\t{clock}}\right)^2 - g^{ij}(\hat{\boldsymbol{r}})\hat{p}_i \hat{p}_j c^2},
\label{eqn:Equant}
\end{align}
with a Weyl's symmetric ordering assumed for the position and momentum operators $\hat{\boldsymbol{r}}$ and $\hat{\boldsymbol{p}}$ that act in the Hilbert space $\mathcal{H}_{\t{cm}}$. Other ordering is also possible, for example by introducing the order parameter $\lambda$, according to the definition $:\!\hat{p}\hat{x}\!:~  \equiv \lambda\hat{p}\hat{x}+(1-\lambda)\hat{x}\hat{p}$, which will be discussed later.

In a further Section, we will consider a general case when both kinematic and gravitational effects are relevant. In the present one, we specialize to the kinematic time dilation, for which the gravitational field is absent and we can substitute $g_{00}(\hat{\boldsymbol{r}}) \to 1$ and $g^{ij}(\hat{\boldsymbol{r}}) \to -\delta^{ij}$ into \eqref{eqn:Equant}. We will investigate whether the quantum time dilation is a universal effect. The notion of universality for classical clocks corresponds to the fact that the magnitude of time dilation does not depend on the mechanism upon which a clock is built. In the case of quantum clocks, the universality of time dilation is defined as the independence of that effect of the Hilbert space $\mathcal{H}_{\t{clock}}$ or the internal Hamiltonian $\hat{H}_\t{clock}$ of the clock.

In what follows, we will only consider the interaction part of dynamics, as local terms are irrelevant for the discussion of time-dilation effects and can always be effectively removed by going to the interaction picture description. Consider the clock prepared in the following initial state:
\begin{equation}
\ket{\Psi(0)} = \int\t{d}^3 \boldsymbol{p}\, \psi(\boldsymbol{p}) \ket{\boldsymbol{p}} \otimes \ket{0},
\label{initial}
\end{equation}
where $\ket{\boldsymbol{p}}\in\mathcal{H}_{\t{cm}}$ represents a well-defined momentum state, $\psi(\boldsymbol{p})$ is the momentum representation of the normalized wave function and $\ket{0}\in\mathcal{H}_{\t{clock}}$ is some initial internal state of the clock.
The expanded Hamiltonian \eqref{eqn:Equant} will contain terms $\propto\frac{\hat{p}^2}{m^2c^2}\hat{H}_{\t{clock}}$ (as we explicitly show in a further Section) that will lead to a momentum-dependent time dilation and the resulting evolution will entangle both degrees of freedom:
\begin{equation}
\ket{\Psi(t)} = \int\t{d}^3 \boldsymbol{p}\, \psi(\boldsymbol{p}) \ket{\boldsymbol{p}} \otimes \ket{\phi(\boldsymbol{p},t)},
\end{equation}
where all the time dependence has been absorbed into $\ket{\phi(\boldsymbol{p},t)}$, which is the evolved internal state of the clock associated with the motional state $\ket{\boldsymbol{p}}$.
The parameter $t$ is the physical time governing the evolution of the system.
However, we will be interested in the operational definition of time corresponding to the measurement outcome $\tau$ performed on the clock's internal degrees of freedom. Let us denote the set of measurement operators providing the time readout $\tau$  as $\{\hat{E}(\tau)\}$. Since the measurement is performed on the internal degrees of freedom only, we can trace out the center-of-mass degrees of freedom and use the reduced density matrix $\hat{\rho}_{\t{clock}}$ representing the internal state of the clock:
\begin{equation}
\label{densitymat}
\hat{\rho}_{\t{clock}}(t) = \int\t{d}^3\boldsymbol{p}\, |\psi(\boldsymbol{p})|^2 \ket{\phi(\boldsymbol{p},t)}\bra{\phi(\boldsymbol{p},t)}.
\end{equation}
Thus the probability ${\cal P}(\tau)$ to obtain the measurement result $\tau$ reads:
\begin{align}
{\cal P}(\tau) &= \t{Tr}\left(\hat{E}(\tau) \hat{\rho}_{\t{clock}}(t) \right) \nonumber \\
&= \int\t{d}^3\boldsymbol{p}\,  |\psi(\boldsymbol{p})|^2 \bra{\phi(\boldsymbol{p},t)} \hat{E}(\tau) \ket{\phi(\boldsymbol{p},t)}.
\label{eqn:marginalprobability}
\end{align}
It is clear from this expression that the phase of the momentum representation wave function $\psi(\boldsymbol{p})$ does not influence the readouts of the clock. Only the probability density of the wave function matters. Therefore, instead of using the initial state of the system given by \eqref{initial}, we could have used a density matrix involving a classical mixture of definite momentum states:
\begin{align}
\hat{\rho}(0) = \int\t{d}^3 \boldsymbol{p}\, |\psi(\boldsymbol{p})|^2 \ket{\boldsymbol{p}}\bra{\boldsymbol{p}} \otimes \ket{0}\bra{0},
\label{rhoclass}
\end{align}
which is a state with erased coherences between different momentum eigenstates in~\eqref{initial}, and the result \eqref{eqn:marginalprobability} would remain identical. In other words, the effective time dilation of the clock is equal to a weighted average, with weights $\abs{\psi(\boldsymbol{p})}^2$ of time dilations of clocks characterized by well-defined momenta. These classical time dilations are universal, i.e., independent of the clock mechanism, therefore the weighted average \eqref{eqn:marginalprobability} of universal time dilations must also be universal. This argument completes the proof of the universality of kinematic quantum time dilation for the case of the clock of an arbitrary quantum state of the clock $\psi(\boldsymbol{p})$.

\textbf{Quantum time dilation for classical states}---Our elementary analysis can shed new light on the interpretation of the quantum time dilation studied in \cite{Smith2020,Grochowski2021}. Let us again notice that the result \eqref{eqn:marginalprobability} can be obtained using the state \eqref{rhoclass} that is a classical mixture of definite momentum states and therefore can be considered completely classical.  This fact raises a question in what sense the \textit{quantum} time dilation based on expression \eqref{eqn:marginalprobability} is and studied earlier in \cite{Smith2020, Grochowski2021} is \textit{quantum}. In order to understand it better, let us revisit the reasoning used in these works.

Let $\ket{\psi_1}$ and $\ket{\psi_2}\in\mathcal{H}_{\mathrm{cm}}$ be two different motional states of the clock, e.g., two Gaussian wave packets representing a clock traveling at two different average speeds. Consider now an initial clock state to be a coherent superposition of the form:
\begin{equation}
\label{twopackets}
\ket{\psi} = \mathcal{N}\left(\cos\theta \ket{\psi_1} + \sin\theta e^{i \varphi} \ket{\psi_2}\right)\in\mathcal{H}_{\mathrm{cm}},
\end{equation}
where $\theta$ represents the respective weights, while $\phi$ is a relative phase between the superposed states. Note that we have introduced a normalization factor $\mathcal{N}$, which is necessary if the superposed states were nonorthogonal.

The momentum representation wave function $\psi(\boldsymbol{p})  \equiv \langle\boldsymbol{p}\ket{\psi}$ of the state \eqref{twopackets} reads:
\begin{equation}
\label{quantum}
\psi(\boldsymbol{p}) = \mathcal{N} \left( \cos\theta \,\psi_1(\boldsymbol{p}) +  \sin \theta e^{i \varphi}\, \psi_2(\boldsymbol{p}) \right)
\end{equation}
and if $\psi_1(\boldsymbol{p})$ and $\psi_2(\boldsymbol{p})$ are nonorthogonal then clearly the variations of the phase $\varphi$ will change the corresponding momentum density distribution $|\psi(\boldsymbol{p})|^2$ that appears in \eqref{eqn:marginalprobability}. As a result, the effective time dilation will indeed depend on the phase $\varphi$. This dependence of the result \eqref{eqn:marginalprobability} on the relative phase $\varphi$ is argued to be the signature of the quantumness of the observed time dilation effect.

In order to quantify the magnitude of the effect, the authors of~\cite{Smith2020,Grochowski2021} compute the difference between the clock rate evaluated for the initial superposed state $\ket{\psi}\otimes\ket{0}$ and its classical counterpart, which is evaluated for a classical mixture of the two wave packets $\ket{\psi_1}$ and $\ket{\psi_2}$:
\begin{equation}
\label{twopacketsclass}
\hat{\rho}(0) = {\cal N}' \left( \cos^2\theta \ket{\psi_1}\bra{\psi_1} + \sin^2\theta \ket{\psi_2}\bra{\psi_2}\right)\otimes\ket{0}\bra{0}.
\end{equation}
That resulting difference quantifies the amount of quantum time dilation. Therefore, this quantum time dilation can be viewed as an effect of state discrimination procedure between a quantum-superposed clock state \eqref{twopackets} and a classical mixture \eqref{twopacketsclass}.

Let us notice that another possibility would be to take the classical counterpart to be \eqref{rhoclass} instead of \eqref{twopacketsclass}, in which case the difference of clock rates would always be zero. This is because the expression \eqref{eqn:marginalprobability} depends only on the diagonal elements of the density matrix in momentum space. Such a construction would produce no quantum contribution to the classical time dilation. Therefore the magnitude of quantum time dilation relies on an arbitrary convention, which classical state should be taken as a reference. 

This ambiguity is the result of the nonorthogonality of the wave packets $\psi_1(\boldsymbol{p})$ and $\psi_2(\boldsymbol{p})$. It could be removed by using orthogonal wave packets, however in this case no quantum time dilation could be observed. Therefore, unlike the double-slit experiment, which reveals the quantum nature of the interference of orthogonal states, quantum time dilation crucially depends on the nonorthogonality of modes $\psi_1(\boldsymbol{p})$ and $\psi_2(\boldsymbol{p})$. Clearly, one may decompose the initial state \eqref{twopackets} in a way, in which the two terms appearing in the decomposition correspond to overlapping wave functions. Hence adding a relative phase between these terms gives rise to a change in momentum distribution appearing in \eqref{eqn:marginalprobability}, but this should be viewed more as an artifact of choosing a particular decomposition of the state rather than a manifestation of a fundamental effect.

Let us go even further and carry out a discrimination procedure between a superposition and a mixture of two states leading to the quantum time dilation, but apply it to an alternative pair of states. The following example illustrates that the nonzero difference can be witnessed even for a pair of completely classical states. Consider the following example of two coherent states $|\alpha\rangle,~|\beta\rangle \in \mathcal{H}_{\mathrm{cm}}$ of the center-of-mass degree of freedom and decompose $|\alpha\rangle$ in a basis $\frac{1}{\sqrt{2}}(|\alpha\rangle\pm|\beta\rangle)$:
\begin{align}
\label{cohsup}
|\alpha \rangle = \frac{1}{2}(|\alpha\rangle + |\beta\rangle) + \frac{1}{2}(|\alpha\rangle - |\beta\rangle).
\end{align}
According to the discrimination procedure discussed above, we should consider a classical counterpart of the state \eqref{cohsup} defined in this case as the following classical mixture:
\begin{align}
\label{cohmix} 
\hat{\varrho}  &\equiv \frac{\cal N}{2}(|\alpha\rangle+|\beta\rangle)(\langle \alpha| + \langle \beta|) 
+\frac{\cal N}{2}(|\alpha\rangle-|\beta\rangle)(\langle \alpha| - \langle \beta|)\nonumber \\
&= {\cal N}(|\alpha\rangle\langle\alpha | + |\beta\rangle\langle\beta |).
\end{align}
In general, both states have different momentum distributions and therefore the measure used to quantify the amount of quantum time dilation for the state \eqref{twopackets} yields a nonzero value for the classical state \eqref{cohsup}.

\textbf{Combined kinematic and gravitational quantum time dilation}---Previously, we analyzed how kinematic time dilation affected the rate of a clock moving along a superposition of trajectories. Our considerations were very general, thanks to the simple structure of the interaction Hamiltonian which distinguished the momentum eigenbasis in $\mathcal{H}_{\mathrm{cm}}$ space and resulted in the clock's reduced density matrix being a mixture of clock states each corresponding to a well-defined momentum of the center-of-mass degree of freedom \eqref{densitymat}. Now we would like to shift gears and consider a combined scenario involving both kinematic and gravitational time dilation. In this case, the interaction Hamiltonian will no longer distinguish a single basis in $\mathcal{H}_{\mathrm{cm}}$.

Let us consider a pointlike clock placed in a Schwarzschild gravitational field at a certain position $\boldsymbol{r}_0$.
The Schwarzschild metric in the proximity of the clock, expressed in terms of the local proper time at the position $\boldsymbol{r}_0$, and proper length $\boldsymbol{r}$ can be written in post-Newtonian approximation derived in Appendix~\ref{Metric}:
\begin{align}
\label{metricexp}
g_{00}(\boldsymbol{r}) &\approx \left(1+\frac{2GM}{r_0 c^2}+\frac{2G^2M^2}{r_0^2c^4}\right)\left(1-\frac{2GM}{rc^2}+\frac{2G^2M^2}{r^2c^4}\right), \nonumber\\ 
g_{ij}(\boldsymbol{r}) & \approx-\delta_{ij}\left(1+\frac{2GM}{rc^2}+\frac{3G^2M^2}{2r^2c^4}\right).
\end{align}
We now substitute the expansion \eqref{metricexp} into the Hamiltonian \eqref{eqn:Equant} and choose the $x$ axis so that it aligns with the local direction of the gravitational field at $\boldsymbol{r}_0$. This leads us to the following approximation to the total Hamiltonian of the clock \eqref{eqn:Equant}:
\begin{align}
\label{decomposition}
\hat{H} \approx \hat{H}_\t{clock}+\hat{H}_\t{cm}(\hat{\boldsymbol{r}}, \hat{\boldsymbol{p}})+\hat{H}_\t{int}.
\end{align}
where the individual terms are obtained by neglecting the higher-order contributions to the total energy, much smaller than the rest energy $mc^2$.
In particular, $\langle\hat{H}_{\t{clock}}\rangle,\abs{ \langle c\hat{p}\rangle},\abs{ \langle mg\hat{x} \rangle},~mg r_0,~\frac{GMm}{r_0}\ll mc^2$. We have:
\begin{align}
\hat{H}_\t{cm} &\equiv mc^2+\frac{\hat{p}^2}{2m}+mg\hat{x}
+\frac{3g}{2mc^2}\overset{\text{\tiny W}}{:}\hat{p}^2\hat{x}\overset{\text{\tiny W}}{:}, 
\nonumber
\\
\hat{H}_\t{int} &\equiv\hat{H}_\t{clock}\underbrace{\left( 
-\frac{\hat{p}^2}{2m^2c^2}
+\frac{g\hat{x}}{c^2}
-\frac{3g}{2m^2c^4}\overset{\text{\tiny W}}{:}\hat{p}^2\hat{x}\overset{\text{\tiny W}}{:}
\right)}_{\hat{V}_1}\nonumber \\ 
&+\hat{H}_\t{clock}^2 \underbrace{
\left(
\frac{\hat{p}^2}{2m^3 c^4}
+\frac{3g}{2m^3c^6}\overset{\text{\tiny W}}{:}\hat{p}^2\hat{x}\overset{\text{\tiny W}}{:}
\right)
}_{\hat{V}_2}
,
\label{int}
\end{align}
where $g\equiv GM/r_0^2$ is the gravitational acceleration at $\boldsymbol{r}_0$, $\hat{x}$ is position operator along the direction of the gravitational field at $\boldsymbol{r}_0$ measuring proper distance from that point, and the symbol $\overset{\text{\tiny W}}{:}\cdot\overset{\text{\tiny W}}{:}$ represents Weyl's symmetric ordering.

In order to characterize the evolution of the system we will use the von Neumann equation for the density matrix $\hat{\rho}^I(t)$ of the total system written in the interaction picture (denoted by the $I$ superscript), in which the interaction Hamiltonian is given by $\hat{H}_\t{int}^I(t) \equiv e^{\frac{i}{\hbar}(\hat{H}_\t{clock}+\hat{H}_\t{cm}) t}\hat{H}_\t{int}e^{-\frac{i}{\hbar}(\hat{H}_\t{clock}+\hat{H}_\t{cm}) t}$:
\begin{align}
i\hbar\pdv{}{t}\hat{\rho}^I(t)=\comm{\hat{H}_\t{int}^I(t)}{\hat{\rho}^I(t)}.
\label{eqn:Neumann}
\end{align}
The approximate solution to Eq.~\eqref{eqn:Neumann} can be found in the first order of the Dyson series by assuming that the initial state of the system exhibits no correlations between the kinematic and internal degrees of freedom: $\hat{\rho}(0) \equiv \hat{\rho}_\t{cm}(0)\otimes\hat{\rho}_\t{clock}(0)$. After going back to the Schr\"{o}dinger picture we can compute the reduced density matrix of the clock's internal state by tracing out the kinematic degrees of freedom $\hat{\rho}_\t{clock}(t) \equiv \Tr_\t{cm}\hat{\rho}(t)$. The details of the calculations are in the Appendix \ref{Evolition} and our final result takes the form:
\begin{align}
\hat{\rho}_\t{clock} (t) &\approx 
\hat{\rho}^0_\t{clock}(t)
-\frac{i}{\hbar}
\Bigg( 
\comm{\hat{H}_{\t{clock}}}{\hat{\rho}^0_\t{clock}(t)}
\Tr
\left(
\hat{V}_1\hat{\rho}_{\t{cm}}(0)
\right) \nonumber \\
&+
\comm{\hat{H}_{\t{clock}}^2}{\hat{\rho}^0_\t{clock}(t)}
\Tr
\left(
\hat{V}_2\hat{\rho}_{\t{cm}}(0)
\right)
\Bigg)
t
\label{eqn:rho1}.
\end{align}
Here $\hat{\rho}^0_\t{clock}(t)\equiv e^{-\frac{i}{\hbar}\hat{H}_{\t{clock}} t}\hat{\rho}_{\t{clock}}(0)e^{\frac{i}{\hbar}\hat{H}_{\t{clock}} t}$ characterizes the evolution of the clock's internal degrees of freedom in the absence of coupling with the kinematic degrees of freedom. Let us compare \eqref{eqn:rho1} with the previously obtained special case \eqref{densitymat} involving only the kinematic time dilation. In that special case, we were able to express the result as a classical mixture of final states corresponding to definite values of momentum. Analogously, for the purely gravitational time dilation without kinematic effects, the result \eqref{densitymatgr} involves a mixture of final states corresponding to definite positions of the clock. Our general result \eqref{eqn:rho1} goes beyond that. Notice that the terms involving the operator $\hat{V}_1$ contain elements that mix up position and momentum operators $\frac{3g}{2m^2c^4}\overset{\text{\tiny W}}{:}\hat{p}^2\hat{x}\overset{\text{\tiny W}}{:}$, see \eqref{int}. For that reason, our general result \eqref{eqn:rho1} cannot be reproduced using a classical mixture of definite position or momentum final states. The above-mentioned terms appearing in \eqref{eqn:rho1} have the form
\begin{align}
&\Tr\left[\overset{\text{\tiny W}}{:}\hat{p}^2\hat{x}\overset{\text{\tiny W}}{:}\hat{\rho}_{\t{cm}}(0)\right]
=\frac{1}{3}\Tr\left[\left(\hat{p}^2\hat{x}+\hat{p}\hat{x}\hat{p}+\hat{x}\hat{p}^2\right)\hat{\rho}_{\t{cm}}(0)\right]
\nonumber \\
&=
\int\dd p \dd x\dd x'~\frac{3xp^2-i\hbar p}{6\pi\hbar}e^{-\frac{i}{\hbar}p(x-x')}\langle x| \hat{\rho}_\t{cm}(0) |x'\rangle.
\label{eqn:nondiagonal}
\end{align}
This result shows that including higher-order terms allows us to find a relativistic correction to the evolution of a clock that depends on combined distributions of position and momentum operators. In this sense, our result exceeds the one presented in the previous Sections, because in general there is no analog of the classical state \eqref{rhoclass} that could produce nonclassical terms of the form \eqref{eqn:nondiagonal}---these terms depend on both position and momenta that do not commute and hence we cannot replace quantum averaging with a classical weighted averaging over states with well-defined positions and momentum. Notice that the result \eqref{eqn:nondiagonal} depends on the type of ordering used in the Hamiltonian \eqref{int}. Ordering other than Weyl's ordering will result in a quantitatively different outcomes. Therefore direct measurements of these higher-order terms can serve as a tool to verify types of orderings realized in specific physical systems.

Our qualitatively new result has not been reported in the literature concerning quantum effects due to gravitational time dilation \cite{Khandelwal2020,Paczos2022,Hadi2022}. To obtain it we had to go beyond second-order perturbation theory and investigate higher-order terms, included in \eqref{eqn:rho1}. The existence of the term proportional to \eqref{eqn:nondiagonal} has no classical counterpart and therefore it can be considered a genuinely quantum contribution.  Finally, this shows that including both kinematic and gravitational degrees of freedom leads to new quantum effects that do not rely on a specific discrimination procedure. It emphasizes the role of noncommutativity of the position and momentum operators.

\begin{figure}[h!]
\centering
    \includegraphics[width=1\linewidth]{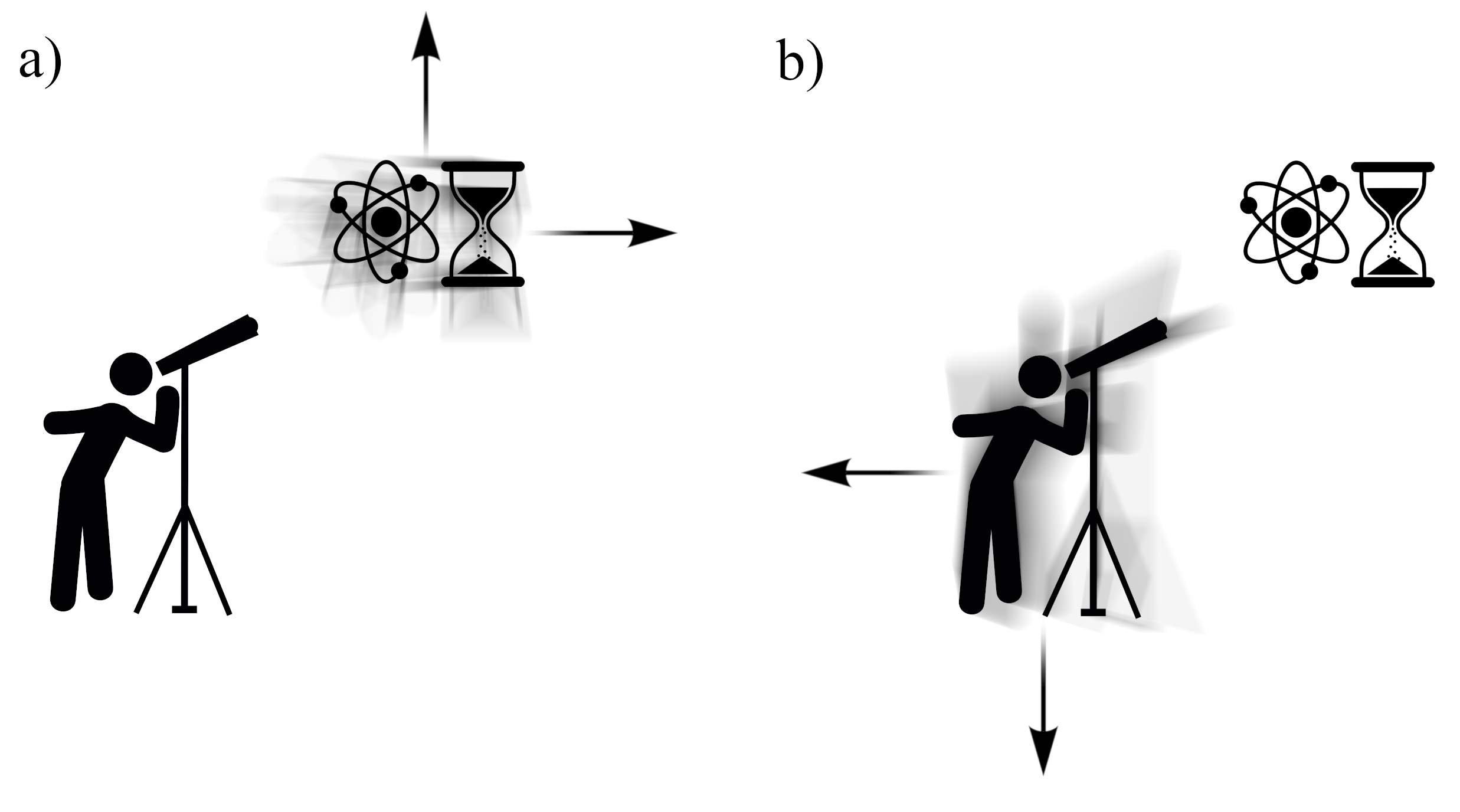}
        \caption{a) A pair of arbitrary clocks moving together along a superposition of two different trajectories, observed from a classical inertial frame; b) a pair of resting clocks observed from a quantum reference frame moving at a superposition of two different velocities.}
        \label{plot}
\end{figure}

\textbf{Nonuniversality of gravitational quantum time dilation}---Let us now consider a scheme involving gravitational time dilation only.
To that end we will use the expansion \eqref{int} in the limit of high values of $m$, for which the kinematic degrees of freedom become less important and which will result in the dropping of the terms depending on momentum:
\begin{align}
\hat{H}_\t{cm} \approx mc^2
+mg\hat{x}
\t{~~~~~~and~~~~~~}
\hat{H}_\t{int} \approx\hat{H}_\t{clock} \frac{g\hat{x}}{c^2}.
\label{int2}
\end{align}
This procedure leads to a gravitational version of \eqref{densitymat}, in which the role of momentum $\hat{\boldsymbol{p}}$ is played by the position $\hat{\boldsymbol{r}}$ of the clock resting in a gravitational field:
\begin{equation}
\label{densitymatgr}
\hat{\rho}_{\t{clock}}'(t) = \int\t{d}^3\boldsymbol{r}\, |\psi'(\boldsymbol{r})|^2 \ket{\phi'(\boldsymbol{r},t)}\bra{\phi'(\boldsymbol{r},t)},
\end{equation}
where $\psi'(\boldsymbol{r})\equiv\braket{\boldsymbol{r}}{\psi}$. This implies that the effective time dilation is equal to the classical weighted average of the time dilations measured by  clocks distributed in space according to the probability distribution $|\psi'(\boldsymbol{r})|^2$. It may seem that the analogy between \eqref{densitymat} and \eqref{densitymatgr} and thus the kinematic and gravitational time dilation is full and therefore the gravitational quantum time dilation should also be universal. However, such an assessment is incorrect. An important implicit assumption we have made in our derivation is that the clock Hamiltonian $\hat{H}_\t{clock}$ appearing in \eqref{decomposition} does not depend on $\hat{\boldsymbol{p}}$ and $\hat{\boldsymbol{r}}$. The independence of $\hat{\boldsymbol{p}}$ is a straightforward consequence of the Lorentz invariance of the theory because the clock's Hamiltonian should be identical in all inertial frames. However, the independence of $\hat{\boldsymbol{r}}$, which embodies the fact that the clock's mechanism is not affected by the strength of the gravitational field, is in fact not true. It is known that every physical system reacts differently to proper acceleration: a pendulum clock and an atomic clock will certainly be affected differently by acceleration. It was shown that no ``ideal clock'' that is completely insensitive to proper acceleration may exist \cite{Lorek2015,EISELE1987,Pierini2018}. The same must also apply to the effect of gravity on the clock's rate. For that reason, the rate of the pendulum clock or any other time-measuring device does depend on the strength of the gravitational field, which has nothing to do with relativity. Therefore even the classical gravitational time dilation cannot be universal not to mention its quantum-superposed counterpart. And the fundamental reason for this fact can be traced back to the argument put forth at the beginning of this paper: only motion with constant velocity is relative. Accelerating is absolute not only relative, and the principle of relativity does not apply to accelerated motions. Therefore the effect of acceleration or gravity on the clock's rate must fundamentally depend on the mechanism of the clock.

\textbf{Conclusions}---Recently studied quantum time dilation effects, both kinematic and gravitational, have risen several issues, including their universality and quantum nature. We have shown that while kinematic quantum time dilation is universal, the gravitational one is not. Our proof of universality provides a necessary condition for quantum reference frames to be well-defined \cite{Angelo2011}. Let us consider a scheme presented in Fig.~\ref{plot}a) in which a pair of arbitrary clocks moving along a superposition of two different trajectories is observed from a classical inertial frame. In Fig.~\ref{plot}b) we show an analogous scheme, in which the pair of clocks is resting, but they are observed from a quantum reference frame moving away at a superposition of velocities. In the latter case, both clocks must be affected by the time dilation exactly the same way. Therefore if both schemes a) and b) are to be equivalent, then the quantum time dilation observed in case a) must be universal.

We also argued that the definition of quantum time dilation relies on an arbitrary choice of the classical reference state and for other, equally justified choices, the effect vanishes. However, we also derived an alternative quantum time dilation effect that manifests itself with higher-order coupling terms between translational and internal degrees of freedom. In this case, the quantum noncommutativity of the involved position and momentum operators guarantees that the effect has no classical analog. Our findings clarify the current understanding of quantum time dilation, providing statements about its universality and underlying nature.

\textbf{Acknowledgments}---We are grateful to Alexander Smith for very insightful discussions and critical comments on the manuscript. K.~D. is financially supported by the (Polish) National Science Center Grant 2021/41/N/ST2/01901. P.~T.~G. is supported by the Foundation for Polish Science (FNP). Center for Theoretical Physics of the Polish Academy of Sciences is a member of the National Laboratory of Atomic, Molecular and Optical Physics (KL FAMO).

\normalem
\bibliography{main}

\onecolumngrid
\newpage
\appendix
\section{Post-Newtonian approximation}\label{Metric}
Let us consider a standard Schwarzschild metric written in Schwarzschild coordinates corresponding to a set of stationary observers, where $t$ is the proper time at infinity \cite{Dragan2021}:
\begin{align}
\dd s^2=\left(1-\frac{2GM}{rc^2}\right) c^2\dd t^2-\left(1-\frac{2GM}{rc^2}\right)^{-1}\dd r^2-r^2 \dd\Omega^2,
\end{align}
and $\dd\Omega^2=\dd \theta^2+\sin^2\theta\dd\phi^2$. Let us reparameterize the radial coordinate according to $r(\varrho)=\varrho\left(1+\frac{GM}{2\varrho c^2}\right)^2$ so that the resulting metric has an isotropic spacial component:
\begin{align}
\dd s^2=\underbrace{\left(\frac{1-\frac{GM}{2\varrho c^2}}{1+\frac{GM}{2\varrho c^2}}\right)^2}_{g_{00}(\varrho)} c^2\dd t^2-\left(1+\frac{GM}{2\varrho c^2}\right)^4\left(\dd \varrho^2+\varrho^2\dd\Omega^2\right).
\label{eqn:Metric1}
\end{align}
In these coordinates the spacial part $(\varrho, \theta, \phi)$ is proportional to the Euclidean metric. Let us now introduce a reparameterized temporal coordinate $\tau_0$ that corrsponds to the proper time measured by a stationary ideal clock placed at $\varrho_0$. We have $\dd \tau_0=\sqrt{g_{00}(\varrho_0)}\dd t$ and the metric in these new coordinates takes the form:
\begin{align}
    \dd s^2=
    \left(\frac{1-\frac{GM}{2\varrho c^2}}{1+\frac{GM}{2\varrho c^2}}\right)^2
 \left(\frac{1-\frac{GM}{2\varrho_0 c^2}}{1+\frac{GM}{2\varrho_0 c^2}}\right)^{-2}
 c^2\dd\tau_0^2-\left(1+\frac{GM}{2\varrho c^2}\right)^4\left(\dd \varrho^2+\varrho^2\dd\Omega^2\right).
\end{align}
Let us proceed with the post-Newtonian expansion of the metric, in which we expand it assuming that the coefficients $GM/\varrho c^2$ and $GM/\varrho_0 c^2$ are small:
\begin{align}
   \dd s^2&\approx
   \left(1-2\frac{GM}{\varrho c^2}+2\left(\frac{GM}{\varrho c^2}\right)^2\right)
   \left(1+2\frac{GM}{\varrho_0 c^2}+2\left(\frac{GM}{\varrho_0 c^2}\right)^2\right)
   c^2\dd\tau_0^2
   \nonumber
   \\
   &-\left(1+2\frac{GM}{\varrho c^2}+\frac{3}{2}\left(\frac{GM}{\varrho c^2}\right)^2\right)
   \left(\dd \varrho^2+\varrho^2\dd\Omega^2\right).
   \label{postnewtonianexpansion}
\end{align}
This brings us to the approximate form of the metric \eqref{metricexp} used in our study.

\section{Evolution of the density matrix}\label{Evolition}
The approximate solution to Eq.~\eqref{eqn:Neumann} can be found in the first order of the Dyson series:
\begin{align}
\hat{\rho}^I(t)\approx\hat{\rho}(0)-\frac{i}{\hbar}\int_{0}^{t}\dd t' \comm{\hat{H}_\t{int}^I(t')}{\hat{\rho}(0)}.
\end{align}
After going back to the Schr\"{o}dinger picture in which $\hat{\rho}(t)\equiv e^{-\frac{i}{\hbar}(\hat{H}_\t{clock}+\hat{H}_\t{cm}) t}\hat{\rho}^I(t) e^{\frac{i}{\hbar}(\hat{H}_\t{clock}+\hat{H}_\t{cm}) t}$ we can compute the reduced density matrix of the clock's internal state by tracing out the kinematic degrees of freedom $\hat{\rho}_\t{clock}(t) \equiv \Tr_\t{cm}\hat{\rho}(t)$:

\begin{align}
    \hat{\rho}_{\t{clock}}(t)\approx e^{-i\hat{H}_{\t{clock}} t}\hat{\rho}_{\t{clock}}(0)e^{i\hat{H}_{\t{clock}} t}
    -\frac{i}{\hbar}\int_{0}^{t}\dd t'\Tr_{\t{cm}}\left(e^{-i\left(\hat{H}_{\t{clock}}+\hat{H}_{\t{cm}}\right) t} \comm{\hat{H}_{\t{int}}^I(t')}{\hat{\rho}(0)}e^{i\left(\hat{H}_{\t{clock}}+\hat{H}_{\t{cm}}\right) t}\right).
    \label{eqn:rhoc}
\end{align}
The first term of \eqref{eqn:rhoc}, $e^{-\frac{i}{\hbar}\hat{H}_{\t{clock}} t}\hat{\rho}_{\t{clock}}(0)e^{\frac{i}{\hbar}\hat{H}_{\t{clock}} t}$,  characterizes the evolution of the internal degrees of freedom of the clock in the absence of coupling with the kinematic degrees of freedom. We will denote this term with $\hat{\rho}^0_\t{clock}(t)$. In order to evaluate the second term we will assume that the initial state of the system exhibits no correlations between the kinematic and internal degrees of freedom: $\hat{\rho}(0) \equiv \hat{\rho}_\t{cm}(0)\otimes\hat{\rho}_\t{clock}(0)$. A relation between the interaction picture and the Schr\"{o}dinger picture for the interaction Hamiltonian \eqref{int} is given by:
\begin{align}
\label{intham}
\hat{H}_\t{int}^I(t') 
= e^{\frac{i}{\hbar}(\hat{H}_\t{clock}+\hat{H}_\t{cm}) t'}
\left(
\hat{H}_\t{clock}\hat{V}_1+
\hat{H}_\t{clock}^2\hat{V}_2
\right)
e^{-\frac{i}{\hbar}(\hat{H}_\t{clock}+\hat{H}_\t{cm}) t'}.
\end{align}
Plugging \eqref{intham} into \eqref{eqn:rhoc} and using the fact that $\comm{\hat{H}_{\t{clock}}}{\hat{H}_{\t{cm}}}=0$ allows us to rewrite \eqref{eqn:rhoc} in the following way:
\begin{align}
    \hat{\rho}_{\t{clock}}(t)\approx 
    \hat{\rho}^0_\t{clock}(t)
    &-\frac{i}{\hbar}\comm{\hat{H}_{\t{clock}}}{\hat{\rho}^0_\t{clock}(t)}
    \int_{0}^{t}\dd t'
    \Tr_{\t{cm}}\left(
    e^{\frac{i}{\hbar}\hat{H}_{\t{cm}} t'}
    \hat{V}_1
    e^{-\frac{i}{\hbar}\hat{H}_{\t{cm}} t'}
    \hat{\rho}_{\t{cm}}(0)
    \right)
    \nonumber
    \\
    &-\frac{i}{\hbar}\comm{\hat{H}_{\t{clock}}^2}{\hat{\rho}^0_\t{clock}(t)}
    \int_{0}^{t}\dd t'
    \Tr_{\t{cm}}\left(
    e^{\frac{i}{\hbar}\hat{H}_{\t{cm}} t'}
    \hat{V}_2
    e^{-\frac{i}{\hbar}\hat{H}_{\t{cm}} t'}
    \hat{\rho}_{\t{cm}}(0)
    \right).
    \label{rhoint}
\end{align}
Let us now employ the Baker–Campbell–Hausdorff formula $e^{\hat{X}}\hat{Y}e^{-\hat{X}}=\hat{Y}+\comm{\hat{X}}{\hat{Y}}+\ldots$ in order to find the exact form of the operators appearing in the trace from \eqref{rhoint}:
\begin{align}
\label{BCH}
e^{\frac{i}{\hbar}\hat{H}_{\t{cm}} t'}
\hat{V}_1
e^{-\frac{i}{\hbar}\hat{H}_{\t{cm}} t'}
&\approx
\hat{V}_1
+
\frac{i}{\hbar}t'
\comm{\hat{H}_{\t{cm}}}{\hat{V}_1},
\\ \nonumber
e^{\frac{i}{\hbar}\hat{H}_{\t{cm}} t'}
\hat{V}_2
e^{-\frac{i}{\hbar}\hat{H}_{\t{cm}} t'}
&\approx
\hat{V}_2
+
\frac{i}{\hbar}t'
\comm{\hat{H}_{\t{cm}}}{\hat{V}_2},
\end{align}
where we neglect quadratic contributions in $t'$ irrelevant in the short-time limit considered in this section. Substituting the result \eqref{BCH} into \eqref{rhoint} finally leads us to \eqref{eqn:rho1}.

\end{document}